# Hot electron transport in suspended multilayer graphene

*Sungbae Lee, Nelka Wijesinghe, Carlos Diaz-Pinto, and Haibing Peng\**

Department of Physics and the Texas Center for Superconductivity, University of Houston, Houston, Texas 77204

\* Corresponding author: haibingpeng@uh.edu

ABSTRACT

We study hot electron transport in short-channel suspended multilayer graphene devices created by a distinct experimental approach. For devices with semi-transparent contact barriers, a dip of differential conductance ($dI/dV$) has been observed at source drain bias $V_d = 0$, along with anomalies at higher $V_d$ likely induced by optical phonon scattering. For devices with low contact barriers, only the $dI/dV$ dip at $V_d = 0$ is observed, and we find a well-fit logarithmic dependence of $dI/dV$ on both the bias $V_d$ and the temperature $T$. The logarithmic $V_d$ dependence is explained with the hot electron effect and the logarithmic $T$ dependence could be attributed to the weak-localization in two-dimensions.

Single and multilayer graphene have attracted enormous interest owing to their unique electronic behaviors [1-3]. The two sublattices of the graphitic structure lead to a chiral electronic behavior, [3] distinct to the semiconductor inversion layers. For single-layer graphene, the low-energy electronic dispersion is linear near the Dirac points **K** and **K'**. For bilayer graphene, the Bernal stacking and the interlayer coupling result in an energy spectrum consisting of four hyperbolic bands, with two bands touching at the Dirac points [4-5]. In multilayer graphene, as more complicated interlayer coupling comes into play, theory predicts different band structures depending on the number of layers and the stacking details [6-8]. Yet experimental investigation is far behind to fuel a better understanding of multilayers [3,9].

To understand the electron transport in graphene flakes, the central point is the scattering mechanism of charge carriers. Extrinsic scattering sources include charge impurities near the graphene lattice [10], microscopic ripples [11], and topological lattice defects [3]. A major intrinsic scattering source is the acoustic phonon, which is suggested to contribute weakly in single-layer graphene [11, 12-13]. Due to the lack of control in scattering sources, the temperature dependence of conductance in graphene flakes is still under debate, and varies from sample to sample [3, 10-13]. In addition, most experiments to date have focused on the transport near zero source-drain bias [14]. Here we report the creation of suspended multilayer graphene devices with the conduction channel a few hundred nanometers in length, and investigate the hot electron transport as a function of source drain bias $V_d$ and temperature $T$. For devices with considerable contact barriers (semi-transparent but not in the tunneling regime), a shallow dip of differential conductance ($dI/dV$) has been observed at $V_d = 0$, along with $dI/dV$ anomalies at higher $V_d$ likely induced by optical phonon scattering. For devices with low contact barriers, only the $dI/dV$ dip at $V_d = 0$ is observed, and we find a well-fit logarithmic dependence of $dI/dV$ on both the bias $V_d$ and the temperature $T$. The logarithmic $V_d$ dependence is explained by the hot electron effect, and the logarithmic $T$ dependence could be attributed to the weak-localization in two-dimensional (2D) systems [15]. Through magnetoconductance experiments, signature of weak localization is obtained for suspended multilayer graphene, which agrees well with the theory considering the chiral nature of graphene carriers [16-18].

Experimentally, we constructed three-terminal devices containing suspended graphene flakes (Fig. 1a). Our strategy is to spin-coat graphene flakes from liquid solutions on to pre-patterned electrodes. We start with a degenerately *n*-doped Si wafer with a 200 nm thick thermal oxide layer on top. By e-beam lithography and lift-off process, parallel metal electrode pairs are patterned on the $SiO_2$ layer with a gap ~ 400 nm between the source and drain. Near the device center, the parallel electrodes are designed to be 2 μm in width, 50 μm in length, and ~ 45 nm in thickness (typically made of 15 nm Pd/ 30 nm Cr). Note that the top surface of electrodes is elevated ~ 45 nm above the $SiO_2$ surface, and it can be elevated further by removing part of the $SiO_2$ layer with buffered HF etching. After that, we spin-coat graphene flakes from liquid solutions on to the above substrate with lithographically patterned electrodes. By chance, some graphene flakes will be suspended between the two elevated source and drain electrodes. Subsequent thermal annealing is carried out at a temperature of 300 ºC in Ar gas for ~ 1 hour to improve the contacts between graphene flakes and electrodes.

Fig. 1b presents a SEM image of a device prepared by the above method, showing a micron-size graphene flake suspended over a gap ~ 400 nm between the source and drain electrodes. The relatively narrow gap between elevated metal electrodes enables us to obtain such devices with suspended graphene flakes electrically connecting the source and drain. Another key point to increase the yield of working devices is to prepare liquid solutions containing enough micron-size thin graphene flakes. We have developed a simple method of mechanical grinding of highly-oriented pyrolytic graphite (HOPG) using a mortar and pestle. High-quality HOPG (SPI, Grade 1) was smashed and initially ground with a mortar and pestle for ~ 6 hours. Graphene flakes were then collected and dispersed in Isopropanol, followed by hours of ultra-sonication. This grinding and sonicating process was repeated several times until a uniform dark-gray solution was obtained. After that, the top portion of the solution was drawn into a separate container, repeatedly diluted and sonicated until the final batch of solution is suitable for spin-coating.

Electrical characterization is carried out in a three terminal configuration (Fig. 1a) by using the degenerately doped silicon as the back gate. Here we study suspended graphene multilayers showing a moderate electric field effect [1-2]. Raman scattering reveals that our multilayer graphene flakes are Bernal (AB) stacked showing asymmetric Raman G′ band, consistent with the findings for samples prepared by mechanical exfoliation [19]. The obtained c-axis crystallite size $L_c$ [19] is close to the flake thickness determined by atomic force microscopy, suggesting the single-crystal quality of graphene flakes prepared by our mechanical-grinding method. This rules out a turbostratic stacking order, which presents a symmetric single Lorentzian G′ band with a larger linewidth than that for monolayer graphene [19]. The electric field effect in multilayer graphene is mainly determined by the flake thickness due to the screening of gate-induced charges by the surface layer [2]. The devices reported below demonstrate 10 % to 20 % conductance tuning with a change of 50 V in gate voltage $V_g$, giving an estimated thickness ~ 10 nm. [2] The *dI/dV* is measured by standard lock-in technique with a small excitation voltage at a frequency 503 Hz, superimposed to the DC bias $V_d$. Fig. 2a illustrates the gate tuning of source-drain current $I_d$ for a multilayer graphene device at $T = 4.2$ K, showing a field effect of ~ 15% as $V_g$ is swept. For the gate tuning of *dI/dV* (Fig. 2a inset), reproducible oscillations are observed, which may be related to the Fabry-Perot interference of electron waves confined between the source and drain. Note that oscillations are absent in DC current $I_d$ versus $V_g$ at finite $V_d$ (Fig. 2a), likely due to the phase averaging for electron waves at different energy. Also, the oscillations in *dI/dV* disappear as temperature increases.

Fig. 2b shows *dI/dV* versus $V_d$ at $V_g = 0$. Two notable features are observed. First, a dip of *dI/dV* clearly shows up near $V_d = 0$, the key feature that we will analyze in details later. Second, anomalies are observed at higher energy outside the dip, which are unlikely due to the variation of electron density of states since they are not shifted accordingly as the gate voltage is varied. For example, anomalies of *dI/dV* appear at $V_d \sim$ -135 mV, -48 mV, +52 mV and +105 mV. This could be caused by the inelastic scattering from optical phonons at the **K** point, which are predicted to be strongly coupled to electrons due to the Kohn anomaly [20-21]. The *dI/dV* anomalies appear to be asymmetric with respect to $V_d$, possibly owing to the different scattering strength for electrons and holes. We note that similar high energy anomalies with the electron-hole asymmetry are observed in different devices [22], indicating its potential origin from optical phonon scattering.

For a second type of devices, we find that high energy anomalies are absent, but the $dI/dV$ dip still occurs at $V_d = 0$ (Fig. 3). Usually such devices show higher conductance (or low contact barriers). For the device of Fig. 3, AFM imaging reveals a thickness of 6.5 nm for the graphene flake and a width similar to the length (~ 400 nm), indicating a nominal conductivity per layer close to $e^2/h$.[23] We suggest that the appearance of optical-phonon induced high-energy anomalies is related to the barriers at the contacts for the first type of devices (Fig. 2). With higher barriers at the contacts in such devices (Fig. 2), high-energy electrons can be injected into graphene flakes from electrodes, which have enough energy to excite optical phonons and induce the high-energy anomalies in $dI/dV$. (We note that the contact barriers for the first type of devices are still semi-transparent, instead of in the tunneling regime, since the minimum $dI/dV$ value inside the dip is still significant and does not drop to zero, in contrast to the case of tunneling-barrier devices in which we observe Coulomb blockade.) On the contrary, for the second type of devices with low contact barriers (Fig. 3), the injected electrons do not have enough energy upon entering graphene flakes, while their acceleration inside graphene is limited by scattering, preventing them from reaching the energy threshold needed to excite optical phonons. Therefore, no high-energy anomalies are observed in such devices with low-contact barriers.

To experimentally confirm the role of contact barriers further, we applied a large DC current (0.2 mA) continuously for ~16 hours inside the cryostat to anneal a graphene device showing both high energy anomalies and the $dI/dV$ dip at $V_d = 0$ (Fig. 4). As shown by the data, after the annealing, the high energy anomalies outside the dip are gone, but the center dip is persistent with similar shape and width. The disappearance of high-energy anomalies as a result of annealing is understandable considering the improvement of contacts by the current annealing, consistent with the previous discussion on the role of contact barriers for high-energy anomalies. In contrast, the persistence of the $dI/dV$ dip at $V_d = 0$ before and after the annealing indicates that the dip is not induced by contact barriers.

Now we come to address the physical origin of the $dI/dV$ dip at $V_d = 0$, which has been observed in more than a dozen devices showing moderate gate tunability in conductance. First, we rule out the possibility that the dip is related to the electronic band structure of graphene flakes (e.g., a band overlap near the neutral point), since it is always pinned at $V_d = 0$ and does not shift as a function of gate voltage (Fig. 3b). Also, similar dips centered at $V_d = 0$ appear in devices showing distinct gate tuning behaviors, no matter if the neutral point can be reached within the investigated gate voltage range or not. On the other hand, the $dI/dV$ dip is unlikely induced by contact barriers, in that the magnitude of the dip is too small for tunneling barriers and the temperature dependence of $dI/dV$ at $V_d = 0$ does not follow the activated behavior (i.e., a temperature dependence $\propto \exp(-a/T)$). (Instead, as we will show later, for the device of Fig. 3 with low contact barriers, the $dI/dV$ at $V_d = 0$ shows a weak, logarithmic temperature dependence.) In addition, the aforementioned annealing experiments (Fig. 4) demonstrate the persistence of the $dI/dV$ dip at $V_d = 0$ with similar shape and width before and after the annealing, suggesting that the dip is not caused by contact barriers.

In Fig. 5a, we replot the data of Fig. 3a, with $V_d$ shown in logarithmic scale. Note that at $T = 4.2$ K, $dI/dV$ is proportional to log $V_d$ (black squares). As the temperature is increased, $dI/dV$ approaches the logarithmic behavior at higher $V_d$, but is saturated at low $V_d$. The saturated plateau regime in $V_d$ is widened as the temperature rises. Similarly, $dI/dV$ at zero bias follows a logarithmic dependence on the temperature $T$ (Fig. 5b), while at finite bias $V_d$, plateaus in $dI/dV$ appear at low temperatures. The logarithmic dependence on both $T$ and $V_d$ suggests that the $dI/dV$ dip at $V_d = 0$ is related to certain heating effect on the lattice or on electrons only. One possibility is that the local lattice temperature could be increased due to Joule heating. Below we estimate the effective lattice temperature based on Joule heating and thermal conductive cooling. The reported thermal conductivity $\kappa$ is ~ 5000 W/m·K for monolayer graphene at room temperature,[24] and the acoustic phonon contribution was suggested to dominate $\kappa$. With a phonon contribution proportional to $T^{1.5}$,[25] one can estimate a lower bound of $\kappa$ to be ~ 8.4 W/m·K at $T = 4.2$ K. With that, we obtain a rise of lattice temperature only ~ 60 mK due to Joule heating at $V_d = 10$ mV for the device of Fig. 5. However, the experimental data (Fig. 5) demonstrate that the effective temperature at $V_d = 10$ mV should be equivalent to $T = 30$ K. Therefore, Joule-heating induced lattice temperature change is too small to explain the observed $dI/dV$ dip.

Instead, the dip can be understood by considering the hot electron effect [26-28]. With a relatively weak electron acoustic-phonon coupling at low $V_d$, electrons are almost out of thermal contact with phonons at 4.2 K, and thus will be heated up by the electric field until the energy gain is balanced by the energy loss due to electron-lattice interaction. This leads to an effective electron temperature higher than the lattice temperature. Although the lattice is still in thermal equilibrium with the cryostat, the differential conductance at finite $V_d$ is governed by the effective electron temperature. Therefore, the $dI/dV$ dependence on $V_d$ follows its dependence on the temperature. (Fig. 5) It is remarkable that the hot electron effect is significant even at temperatures near 100 K. This indicates a weak acoustic-phonon scattering, which leads to a weak temperature dependence of electron transport in multilayer graphene. As shown in Fig. 5b, the differential conductance at $V_d = 0$ changes only ~ 10 % from $T = 4.2$ K to 100 K.

The observed logarithmic temperature dependence in the low-contact barrier device (Fig. 5) could be ascribed to the weak localization in 2D systems [15,26]. Considering phonon as the dominant inelastic scattering source, Anderson *et al.* [26] suggested that the ratio of coefficients of $\log V_d$ to that of $\log T$ (i.e. $S_V/S_T$) should be $2/(2+p)$, where the constant $p$ can be 2, 3 or 4 depending on the temperature and the phonon dimension. Fig. 5a shows a fitting to the 4.2 K data by $dI/dV = G_0 + S_V \log V_d$, and Fig. 5b provides a fitting to the data of $V_d = 0$ by $dI/dV = G_1 + S_T \log T$, where $G_0$, $S_V$, $S_T$, $G_1$ are constant fitting parameters. For this device, we obtain $S_V/S_T = 0.94$ from the logarithmic fittings of the data, which is larger than the maximum allowed ratio $S_V/S_T = 1/2$ according to the theory [26]. Out of seven devices (Table 1) with reasonably well-observed logarithmic behaviors, the extracted $S_V/S_T$ has a mean value of 0.97 with a standard deviation of 0.12. (We note that the device of Fig. 2 is not included to extract intrinsic parameters since the considerable contact barriers cause deviations from a good logarithmic behavior.) Unlike metal films [27], our experimental results from graphene flakes could not be explained with the allowable $p$ values for phonon scattering suggested in Ref. 26. It is also different than the hot electron effect in silicon inversion layers [28], where the electron temperature change $\Delta T \propto V_d^{3/2}$ and this should lead to $S_V/S_T = 3/2$. If one considers the electron specific heat linearly dependent on $T$, our result of $S_V/S_T \sim 1$ implies an inelastic scattering time nearly independent of temperature for multilayer graphene [26]. Future theoretical work is needed to quantitatively address the hot electron effect and the inelastic scattering mechanism in graphene systems.

Finally we check the signature of weak localization in multilayer graphene from magnetoconductance experiments. It has been suggested [16-18, 29-32] that weak localization in single-layer and bilayer graphene is different than that in conventional 2D systems [15], owing to their chiral electrons. The weak localization magnitude is not only sensitive to inelastic phase-breaking processes, but also dependent on elastic scattering processes. The inset of Fig. 5b illustrates the change of differential conductance, $\Delta\sigma(B) = dI/dV|_B - dI/dV|_{B=0}$, as a function of magnetic field $B$ perpendicular to the graphene flake. The experimental data clearly demonstrate a positive magnetoconductance. However, the data could not be fitted by the weak localization theory including only the inelastic scattering [15]. Instead, the data can be fitted well by considering both the elastic and inelastic scattering for graphene [16].

$$\Delta\sigma(B) = \frac{e^2}{\pi h}\left[F\left(\frac{B}{B_\phi}\right) - F\left(\frac{B}{B_\phi + 2B_i}\right) - 2F\left(\frac{B}{B_\phi + B_i + B_*}\right)\right]. \quad (1)$$

Here $F(z) = \ln z + \psi(0.5 + z^{-1})$ with $\psi(x)$ the digamma function, and $B_{\phi,i,*} = \hbar/4eL_{\phi,i,*}^2$. $L_\phi$ is the dephasing length, $L_i$ is the intervalley scattering length, and $L_*$ characterizes the trigonal warping effect and chirality-breaking elastic intravalley scattering. We note that Eq. (1) was initially derived for single-layer graphene, but it works well to describe our experiments for multilayer graphene (Fig. 5b inset). This may be explained by the trigonal symmetry in the corner of the hexagonal Brillouin zone for general graphene layers [18]. We thank Dr. V. Hadjiev for Raman scattering and Dr. C.S. Ting for helpful discussions.

FIGURE CAPTIONS

FIG. 1. (Color online) (a) Schematic cross-section (upper panel) and top view (lower panel) of the device structure. (b) SEM image of a device with a suspended graphene flake bridging electrodes. Scale bar: 1μm.

FIG. 2. (a) Source drain current $I_d$ vs. gate voltage $V_g$ under a DC bias $V_d = 100$ mV (Inset: differential conductance $dI/dV$ vs. $V_g$ at $V_d = 0$), and (b) $dI/dV$ vs. $V_d$ with $V_g = 0$ at $T = 4.2$ K for a suspended multilayer graphene device.

FIG. 3. (Color online) (a) Differential conductance $dI/dV$ vs. DC bias $V_d$ with gate voltage $V_g = 0$ at temperatures from 4.2 K to 100 K for another device. (b) $dI/dV$ vs. $V_d$ at temperature $T = 4.2$ K for $V_g = $ - 50 V, 0 V, and 50 V, respectively. (c) $dI/dV$ vs. $V_g$ with $V_d = 0$ at $T = 4.2$ K.

FIG. 4 (Color online) Electron transport data before and after current annealing for another suspended multilayer graphene device showing both high energy anomalies and the $dI/dV$ dip at $V_d = 0$. (a) $dI/dV$ vs. $V_d$ with $V_g = 0$ and (b) $dI/dV$ vs. $V_g$ with $V_d = 0$ at $T = 4.2$ K for the device before current annealing. (c) $dI/dV$ vs. $V_d$ with $V_g = 0$ and (d) $dI/dV$ vs. $V_g$ with $V_d = 0$ at temperatures from 4.2 K to 100 K for the same device after annealed with a DC current of 0.2 mA for 16 hours inside the cryostat. In the data of $dI/dV$ vs. $V_g$, reproducible Fabry-Perot oscillations appear and they are suppressed as temperature increases.

FIG. 5. (Color online) (a) A replot of data from Fig. 3(a) with positive $V_d$ presented in a logarithmic scale (symbols), and a fitting to the 4.2 K data by $dI/dV = G_0 + S_V \log V_d$ with $S_V = 19.6$ μS. (b) The temperature dependence of $dI/dV$ for $V_d = 0$, 10 mV and 20 mV, respectively, and a fitting to the data of $V_d = 0$ by $dI/dV = G_1 + S_T \log T$ with $S_T = 20.8$ μS. Inset: change of differential conductance $\Delta\sigma(B) = dI/dV|_B - dI/dV|_{B=0}$ under a perpendicular magnetic field $B$ at $T = 4.2$ K with $V_d = 0$ (symbols), and the best fitting (line) by Eq. (1) described in text with $L_\phi = 73$ nm, $L_i = 88$ nm, and $L_* = 1.3$ nm.

TABLES

| Samples | $S_V$ (µS) | $S_T$ (µS) | $S_V/S_T$ |
|---|---|---|---|
| #1 | 5.36 | 6.12 | 0.88 |
| #2 | 14.2 | 16.6 | 0.86 |
| #3 | 17.7 | 19.4 | 0.91 |
| #4 | 55.0 | 57.1 | 0.96 |
| #5 | 11.8 | 9.86 | 1.20 |
| #6 | 10.8 | 10.3 | 1.05 |
| #7 | 19.6 | 20.8 | 0.94 |

Table 1. $S_V$, $S_T$ and $S_V/S_T$ values extracted from the experimental data for seven devices.

FIGURES

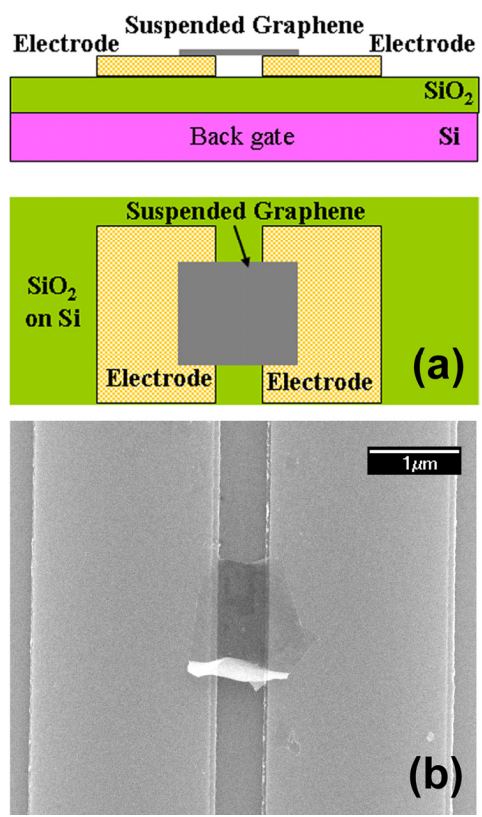

Fig. 1

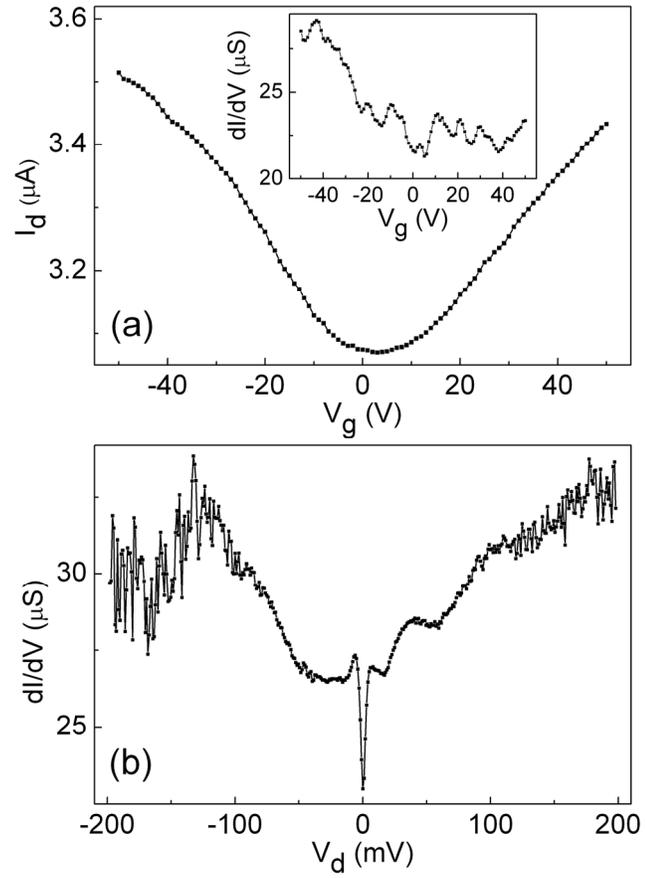

Fig. 2

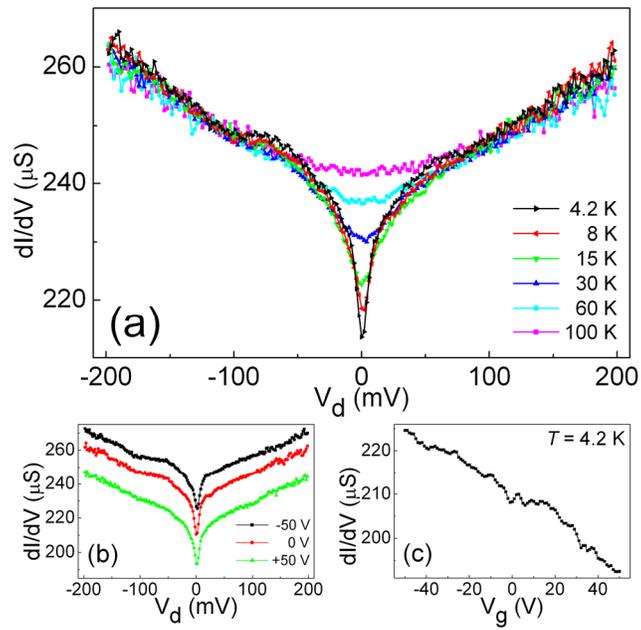

Fig. 3

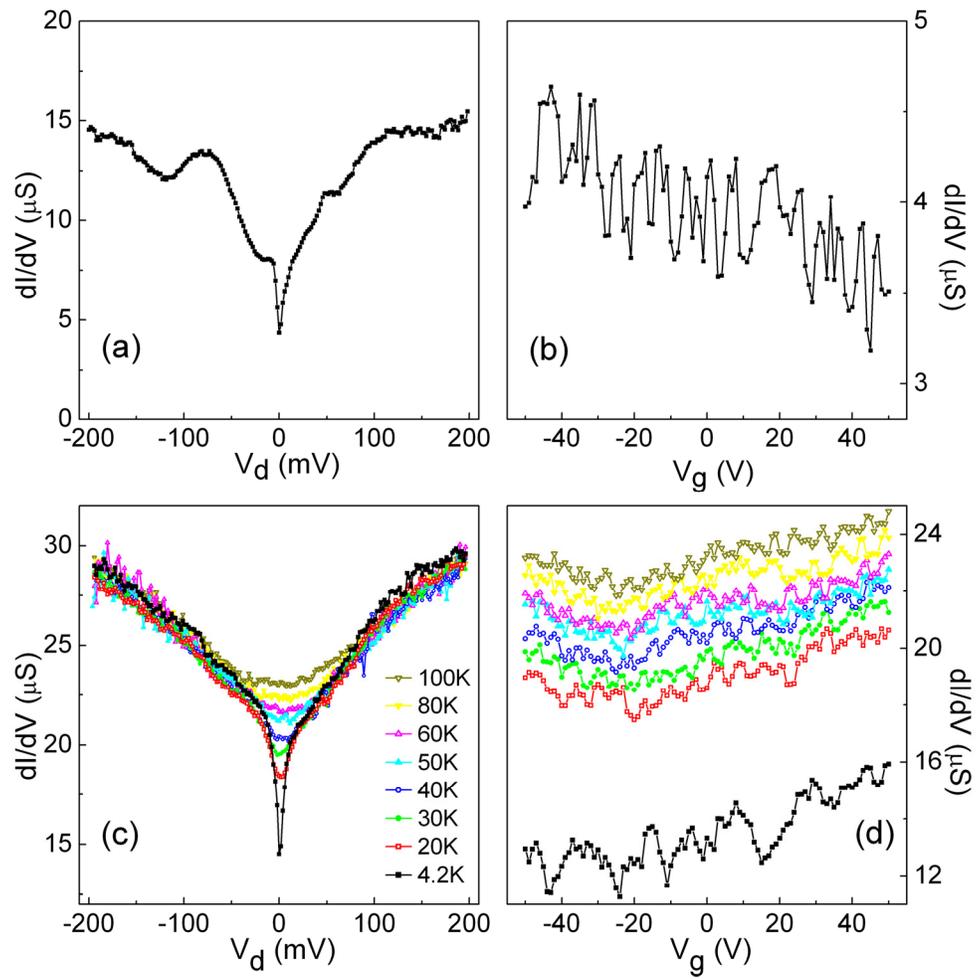

Fig. 4

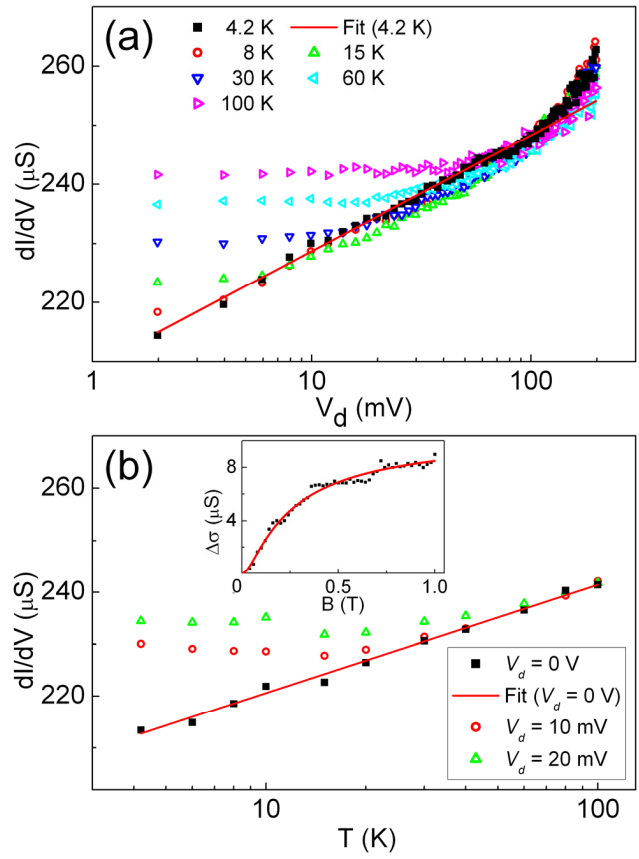

Fig. 5